\documentclass[preprint,eqsecnum,nofootinbib]{revtex4}
\usepackage{epsfig}
\usepackage{graphicx}
\usepackage{dcolumn}
\usepackage{amsmath}
\usepackage{enumerate}

\def\beq{\begin{equation}}
\def\eeq{\end{equation}}

 \def\gtap{\mathrel{ \rlap{\raise 0.511ex \hbox{$>$}}{\lower 0.511ex
   \hbox{$\sim$}}}} 
\def\ltap{\mathrel{ \rlap{\raise 0.511ex
    \hbox{$<$}}{\lower 0.511ex \hbox{$\sim$}}}} 
\newcommand{\bea}{\begin{eqnarray}} \newcommand{\eea}{\end{eqnarray}}

\newcommand{\deltasol}{\mbox{$\Delta m_{21}^2 $}}

\newcommand{\deltaatm}{\mbox{$\Delta m_{31}^2 $}}

\newcommand{\tetaot}{\mbox{$\theta_{13}$}}

\newcommand{\nova}{\mbox{NO$\nu$A}}

\begin{document}

\vskip-6pt \hfill {RM3-TH/08-4} 
\vskip-6pt \hfill {Roma-TH-1465} 
\vskip-6pt \hfill {IPPP/07/100} 
\vskip-6pt \hfill {DCPT/07/200}

\title{An intermediate $\gamma$ beta-beam neutrino experiment \\ with
  long baseline}

\author{
\mbox{Davide Meloni$^1$}, 
\mbox{Olga Mena$^2$}, 
\mbox{Christopher Orme$^3$}, 
\mbox{Sergio Palomares-Ruiz$^3$} 
\mbox{and Silvia Pascoli$^3$}} 

\affiliation{\vspace{2mm} 
\mbox{$^1$ Dipartimento di Fisica, Universit\'a di
  Roma Tre and INFN, Sez. di Roma Tre,}
\mbox{via della Vasca Navale 84, I-00146 Roma, Italy}
\mbox{$^2$ INFN - Sez. di Roma, Dipartimento di Fisica,
  Universit\'a di Roma "La Sapienza",}
\mbox{P.le A. Moro, 5, I-00185 Roma, Italy}
\mbox{$^3$ IPPP, Department of Physics, Durham University,
  Durham DH1 3LE, United Kingdom}}


\begin{abstract}
In order to address some fundamental questions in neutrino physics a
wide, future programme of neutrino oscillation experiments is
currently under discussion. Among those, long baseline experiments
will play a crucial role in providing information on the value of
$\theta_{13}$, the type of neutrino mass ordering and on the value of
the CP-violating phase $\delta$, which enters in 3-neutrino
oscillations. Here, we consider a beta-beam setup with an intermediate
Lorentz factor $\gamma = 450$ and a baseline of 1050~km. This could
be achieved in Europe with a beta-beam sourced at CERN to a detector
located at the Boulby mine in the United Kingdom. We analyse the
physics potential of this setup in detail and study two different
exposures ($1 \times 10^{21}$ and $5 \times
10^{21}$~ions-kton-years). In both cases, we find that the type of  
neutrino mass hierarchy could be determined at 99\% CL, for all values
of $\delta$, for $\sin^2 2\theta_{13} > 0.03$. In the high-exposure
scenario, we find that the value of the CP-violating phase $\delta$
could be measured with a 99\% CL error of $\sim$20$^o$ if $\sin^2
2\theta_{13} > 10^{-3}$, with some sensitivity down to values of
$\sin^2 2\theta_{13} \simeq 10^{-4}$. The ability to determine the
octant of $\theta_{23}$ is also studied, and good prospects are found
for the high-statistics scenario.
\end{abstract}
\maketitle


\section{Introduction}

In recent years, compelling evidence for neutrino oscillations has
been obtained in atmospheric~\cite{atm,SKatm},
solar~\cite{sol,SKsolar,SNO}, reactor~\cite{CHOOZ,PaloVerde,KamLAND}
and long-baseline accelerator~\cite{K2K,MINOS} neutrino
experiments. They have measured with good accuracy the oscillation 
parameters: two mass squared differences, $\Delta m^2_{31}$ and
\deltasol, where $\Delta m^2_{ji} \equiv m_j^2 - m_i^2$, and two mixing
angles, $\theta_{12}$ and $\theta_{23}$. The third mixing angle
$\theta_{13}$ is small and strongly constrained. A combined
analysis of all present data gives for the best fit values and the
$2\sigma$ allowed ranges of the measured oscillation
parameters~\cite{TSchwNuFact07}:
\begin{eqnarray}
\label{deltaatmvalues}
(|\deltaatm|)_{\rm BF} = 2.4 \times 10^{-3} \, \mathrm{eV}^2, 
& \hspace{5mm} 2.1  \times 10^{-3} \ \mathrm{eV}^2 \leq |\deltaatm| 
\leq 2.7 \times 10^{-3} \ \mathrm{eV}^2 ,  \\
\label{deltasolvalues}
(\deltasol)_{\rm BF} = 7.6 \times 10^{-5} \ \mathrm{eV}^2, 
&  \hspace{5mm} 7.3 \times 10^{-5} \ \mathrm{eV}^2 \leq \deltasol \leq
8.1 \times  
10^{-5} \ \mathrm{eV}^2, \\ 
\label{sinatmvalues}
(\sin^2 \theta_{23})_{\rm BF} = 0.50,
& 0.38 \leq \sin^2 \theta_{23} \leq 0.63, \\
\label{sinsolvalues}
(\sin^2 \theta_{12})_{\rm BF} = 0.32,
& 0.28 \leq \sin^2 \theta_{12} \leq 0.37 \,.
\end{eqnarray}
The combined limit on the $\theta_{13}$ mixing angle
reads~\cite{TSchwNuFact07}
\beq
\sin^2\theta_{13} < 0.033~(0.050)
\quad\mbox{at}\quad 2\sigma \, (3\sigma)~.
\label{th13}
\eeq
Despite the remarkable, recent progress in our understanding of
neutrino physics, fundamental questions need to be addressed in the
future in order to shed light on the theory beyond the Standard Model
of particle interactions which is responsible for neutrino masses and
mixing. We need to establish the nature of neutrinos (whether Dirac
or Majorana particles), the neutrino mass ordering (normal or
inverted), the absolute neutrino mass scale, the value of the unknown
mixing angle $\theta_{13}$, and if the CP-symmetry is violated in the
leptonic sector. In addition, it will be important to determine with
better precision the already known oscillation parameters. 
 
A wide, future programme of neutrino oscillation experiments, under
discussion at present, addresses some of the issues mentioned
above~\cite{ISS,Euronu}. In particular, long baseline experiments will
aim at providing information, first on the values of $\theta_{13}$,
and then on the type of neutrino mass ordering and on the value of the
CP-violating phase $\delta$, which enters in 3-neutrino
oscillations. Superbeams, neutrino factory and beta-beams are studied
in detail. Superbeams extend the present experimental concepts for
conventional beams with an upgrade in intensity and in detector
size. Various proposals are under consideration or construction:
T2K~\cite{T2K} in Japan, NO$\nu$A~\cite{newNOvA} in the US, and the
possibility of a wide-band beam if $\theta_{13}$ turns out to be
large~\cite{Barger:2007jq}. Neutrino factories~\cite{nufact} and
beta-beams~\cite{zucchelli,mauro1,mauro2} are novel concepts. In a
neutrino factory, muons (antimuons) are produced, cooled and
accelerated to high Lorentz factor before stored in a decay ring. Muon
neutrino (antineutrino) and electron antineutrino (neutrino) beams are 
produced and aimed at a magnetised detector at very far distance.
Magnetisation is necessary in order to separate the right muon
disappearance signal from the wrong muon appearance signal, which is
sensitive to matter effects and CP violation.
Beta-beams~\cite{zucchelli,mauro1,mauro2} exploit ions which are
accelerated to high Lorentz factors, stored and then $\beta$-decay,
producing a collimated electron neutrino beam. The typical neutrino
energies are in the 200~MeV--GeV range, requiring detectors with
hundred-of-MeV thresholds and good energy resolution. Lower energies
imply shorter baselines and, typically, baselines of few hundred of km
are considered. The only requirement is good muon identification in
order to detect the appearance of muon neutrinos (or antineutrinos)
from the initial electron neutrino (or antineutrino) beam. Hence, in
principle, no magnetisation is required and therefore water
\v{C}erenkov, totally active scintillator, liquid argon detectors and
non-magnetised iron calorimeters could be used, depending on the peak
energy.

As it is well known, the determination of the mixing angle
$\theta_{13}$, the type of neutrino mass hierarchy, if the CP-symmetry
is violated in the leptonic sector and the octant of the mixing angle
$\theta_{23}$, is severely affected by
degeneracies~\cite{deg1,deg2,deg3,deg4,deg5}. For a fixed baseline and
neutrino energy different sets of the unknown parameters
$(\theta_{13}, \delta, \mbox{sgn}(\deltaatm), \theta_{23}\neq \pi/4)$ 
provide an equally good fit to the probability for neutrino and
antineutrino oscillations. Therefore, a measurement of these
probabilities in an experiment, even if very accurate, might not allow
to discriminate between the various allowed solutions. Various
strategies have been envisaged in order to weaken or resolve this
issue: from exploiting the energy dependence of the signal in the same
experiment~\cite{Barger:2007jq}, to the combination of different 
experiments~\cite{otherexp1,otherexp2,otherexp3,otherexp4,otherexpbeta,otherexp5,otherexp6,otherexp6,otherexp7,otherexp8},
to using more than one baseline for the same
beam~\cite{BNLreport,MN97,silver,BMW02off,SN,twodetect,CDFL07}. On the
other hand, the mixing angle $\theta_{13}$ also controls the Earth
matter effects in multi-GeV
atmospheric~\cite{atmmatter1,mantle,core,atmmatter2,atmmatter3} and in
supernova~\cite{SNastro} neutrino oscillations, and the magnitude of
the T-violating and CP-violating terms in neutrino oscillation
probabilities is directly proportional to
$\sin\theta_{13}$~\cite{CPT}. Therefore, the determination of 
$\theta_{13}$ is crucial for the future possibilities at neutrino
oscillation experiments of pinning down the type of mass hierarchy, if
the CP-symmetry is violated in the lepton sector and the octant of
$\theta_{23}$.

In beta-beam experiments, due to the energy dependence of the neutrino
flux and of the relevant cross sections for the interactions in the
detector, in general a better sensitivity to the type of hierarchy and
CP violation can be reached for higher gammas and consequently longer
baselines~\cite{betabeamhigh,bblindner}. In particular, matter effects
increase with distance and energy. Neutrino oscillation experiments
with baselines of few hundred km, as the CERN-Frejus option or the T2K
superbeam, turn out to have no sensitivity to the sign of \deltaatm,
for the allowed values of
$\theta_{13}$~\cite{lowgamma,otherexpbeta,CMMS06,bblindner}. Higher
energy setups for baselines $>500$~km have also been studied in
detail~\cite{betabeamhigh,bblindner,alternating,betaCERNupgrade,bblhc,CDFL07}. 

Equipping the CERN Super Proton Synchrotron (SPS) with a fast cycling
superconducting magnet would provide a fast ramp which would avoid a
significant loss of ions by decay in the accelerating phase and would
allow to reach high gammas. The studies in Ref.~\cite{betaCERNupgrade}
considered the reach of this setup using an iron magnetised detector
located at Gran Sasso. They showed a very good physics reach, using
both neutrinos from $^{18}$Ne and antineutrinos from $^6$He. However,
longer baselines can be considered in Europe. In particular, it has
been recently pointed out that the Boulby mine on the north-east coast
of England has excellent potential for expansions~\cite{Boulby}. This
would allow to excavate laboratories able to host detectors with
mass of a few tens of ktons, as required in a long baseline
experiment. Therefore, Boulby constitutes a very interesting option
for a future long baseline experiment in Europe, allowing for longer
distances from CERN than Frejus, Canfranc and Gran Sasso. In the
present article, we exploit this new opportunity and consider a
neutrino beta-beam sourced at CERN and a detector located in the
Boulby mine. This choice of setup has a baseline of 1050~km that
allows a superior sensitivity to matter effects, as well as to CP
violation with respect to lower energy possibilities. As just
mentioned, our choice is motivated on one side by the possibility of
an upgrade of the accelerator complex at CERN and on the other by the
recent studies at the Boulby mine which indicate the possibility to
build large caverns in hard stable rock at this
site~\cite{Boulby}. Differently from Ref.~\cite{betaCERNupgrade}, we
consider a detector with low energy threshold and good energy
resolution. This has important physics implications as it allows one
to fully exploit the oscillatory pattern of the signal and in
particular to be sensitive to both the first and the second
oscillation maximum. As the dependence of the signal on CP violation
and matter effects is very different at different energies, it is
possible to resolve degeneracies and reach an excellent sensitivity
with a neutrino run only. In fact, sufficient information can be
extracted from the neutrino signal alone and the antineutrino run,
suppressed by small cross sections, does not improve the physics reach
and it has not been included in our study.

The paper is organised as follows. In Section~II we describe the
beta-beam setup and the resulting neutrino flux. In Section III we
discuss the strategy beyond the choice of experimental setup and in
particular we discuss how a neutrino run can resolve degeneracies if
the oscillatory pattern of the probability is fully exploited. In
Sections IV and V we give the details of the numerical analysis and
its results for the physics reach of the setup. Finally, in Section VI
we draw our conclusions.


\section{The beta-beam setup}

First introduced by Zucchelli~\cite{zucchelli}, a beta-beam experiment
exploits a well collimated neutrino beam produced by the acceleration
and subsequent decay of stored $\beta$-emitting ions. The neutrino
flux is very well known since the beta decay is well understood
theoretically and all forward going neutrinos are collimated into a
cone with opening angle $1/\gamma$, $\gamma$ being the ion boost in
the laboratory frame. The dominant factor in the choice of ion is the
need for a high luminosity at the detector site. Potential ions need
to have a small proton number to minimise space charge, and half-lives
$\sim$ 1 second to reduce ion losses during the acceleration while
still maintaining a large amount of useful decays per year. The most
promising candidate ions are $^{18}$Ne and $^{8}$B for neutrinos, and
$^{6}$He and $^{8}$Li for anti-neutrinos. In Table~\ref{T:distances}
we show, for each of these four isotopes, the energy at the peak of
the beta-beam spectrum in the rest frame and the value of this energy
in the boosted frame for the current SPS and for the upgraded SPS. We
also show the baseline for which the peak energy would correspond to
the first oscillation maximum of the probability of transition of
$\nu_e$ into $\nu_\mu$.

\begin{table} 
  \begin{center}
    \begin{tabular}{|c|c|c|c|c|c|c|c|}
\hline 
\hline 
\multicolumn{2}{|c|}{} & \multicolumn{3}{c|}{Current SPS} &
\multicolumn{3}{c|}{Upgraded SPS} \\[1ex] 
\hline
\hline 
Isotope & $E_{\rm P}$ (MeV) & $\gamma$ & $E_{\nu}$ (GeV) & $L_{\rm
  max}$ (km) & $\gamma$ & $E_{\nu}$ (GeV) & $L_{\rm max}$ (km) \\[1ex]
\hline 
\hspace{2mm} $^{18}$Ne \hspace{2mm} & \hspace{2mm} 1.86 \hspace{2mm} &
\hspace{2mm} 270 \hspace{2mm} & \hspace{2mm}  1.0 \hspace{2mm} &
\hspace{2mm} 510 \hspace{2mm} & \hspace{2mm} 590 \hspace{2mm} &
\hspace{2mm} 2.2 \hspace{2mm} & \hspace{2mm} 1130 \hspace{2mm} \\[1ex]
\hspace{2mm} $^{6}$He \hspace{2mm} & \hspace{2mm} 1.94 \hspace{2mm} &
\hspace{2mm} 160 \hspace{2mm} & \hspace{2mm} 0.6 \hspace{2mm} &
\hspace{2mm} 320 \hspace{2mm} & \hspace{2mm} 355 \hspace{2mm} &
\hspace{2mm} 1.4 \hspace{2mm} & \hspace{2mm} 710 \hspace{2mm} \\[1ex]
\hline 
$^{8}$B & 7.37 & 300 & 4.4 & 2270 & 670 & 9.8 & 5060 \\[1ex]
$^{8}$Li & 6.72 & 180 & 2.4 & 1240 & 400 & 5.4 & 2770 \\[1ex]
\hline
\hline
\end{tabular}
\end{center}
\caption{Energy at the peak of the beta-beam spectrum in the rest
  frame ($E_{\rm P}$) and in the boosted frame for the current
  (maximum proton energy of 450~GeV) and upgraded (maximum proton
  energy of 1~TeV) SPS. Also shown the maximum achievable   $\gamma$
  factor in both cases for each isotope. The maximum baseline, $L_{\rm
  max}$, represents the distance at which the first oscillation
  maximum of the probability of conversion of $\nu_e$ into $\nu_\mu$ is
  located.}
\label{T:distances}
\end{table}

In the rest frame of the ion, the electron neutrino flux depends on
the neutrino energy, $E_\nu$, as
\begin{equation}
\frac{d\Phi^{\mathrm{rf}}}{d\cos\theta dE_{\nu}} \sim
E_{\nu}^{2}(E_{0}-E_{\nu})\sqrt{(E_{\nu}-E_{0})^{2}-m_{e}^{2}} ~. 
\label{E:flux}
\end{equation}
Here, $E_{0}$ is the end-point energy of the decay and $m_{e}$ is the
mass of the electron. The neutrino flux per solid angle at the
detector located at distance $L$ from the source after boost $\gamma$
is~\cite{betabeamhigh}
\begin{equation}
\left.\frac{d\Phi^{\mathrm{lab}}}{d\Omega dy}\right|_{\theta\simeq 0}
\simeq \frac{N_{\beta}}{\pi
  L^{2}}
\frac{\gamma^{2}}{g(y_{e})}y^{2}(1-y)\sqrt{(1-y)^{2}-y_{e}^{2}} ~,
\label{E:Bflux}
\end{equation}
where $0 \leq y=\frac{E_{\nu}}{2\gamma E_{0}}\leq 1-y_{e}$,
$y_{e}=m_{e}/E_{0}$, $N_{\beta}$ is the number of useful ion decays
per year, and 
\begin{equation}
g(y_{e})\equiv \frac{1}{60}\left\{
\sqrt{1-y_{e}^{2}}(2-9y_{e}^{2}-8y_{e}^{4})+15y_{e}^{4}
\log\left[\frac{y_{e}}{1-\sqrt{1-y_{e}^{2}}}\right] \right\} ~. 
\end{equation}
A neutrino with energy $E^{\mathrm{rf}}$ in the rest frame will have a
corresponding energy $E_{\nu} = 2\gamma E^{\mathrm{rf}}$ in the
laboratory frame along the $\theta =0^{\circ}$ axis. Consequently,
ions with lower end-point values in the ion rest frame require higher
$\gamma$ in order to achieve the neutrino energies appropriate to a 
given baseline. Most studies consider $^{18}$Ne and $^{6}$He due to
their low Q-values. For similar ion production rates, this choice will
provide a more focused beam and, the flux scaling as $\gamma^{2}$,
higher fluxes at the detector. It is noted also that placing a
detector off-axis still constitutes a Lorentz boost so that, unlike a
superbeam, the spectral shape is maintained although the mean energy
will be lower than in the on-axis case.

The initial study of a beta-beam~\cite{zucchelli} considered a
`low-$\gamma$' machine which experimentally had three stages: nuclide
production via the Isotope Separation OnLine (ISOL) technique;
acceleration using existing technology such as the CERN Proton
Synchrotron (PS) and SPS before storing the ions  in a decay ring. The
feasibility of such scheme has been demonstrated~\cite{feasibility},
the current magnetic rigidity of the SPS allowing a maximum $\gamma
\sim 160$ for $^6$He and $\gamma \sim 270$ for $^{18}$Ne. The ions
will be accelerated to 300 MeV/amu through the use of a linac and
rapid cycling synchrotron before being fed into the CERN PS. It is
envisaged that there will be 16 bunches of 2.5 $\times$ 10$^{12}$ ions
which will be merged to 8 upon acceleration to $\gamma=9$. The final
phase of the acceleration requires the transfer to the SPS where they
will be accelerated to the $\gamma$ required for the experiment. The
ions are then stacked in a decay ring so that enough ions decay to
achieve a useful neutrino flux. For the $\gamma=100$ scenario, it is
proposed to have a decay ring with the same circumference as the SPS
(6880 m) but in a `racetrack' design with 2500 m straight
sections. For a single baseline beta-beam, $\sim 35 \%$ of the
neutrinos will be available  from a single straight section. With the
current SPS, the CERN-Frejus baseline ($L\sim$~130 km) is the only
option available. However, the short distance does not allow
sensitivity to matter effects and to the neutrino mass ordering.

Some LHC upgrade scenarios conceive implementation of the SPS with
fast cycling superconducting magnets leading to the injection of 1 TeV
protons into the LHC. Such a setup would  allow the acceleration of
$^{18}$Ne and $^{6}$He up to $\gamma$ of 580 and 350,
respectively. Various studies have exploited this
possibility~\cite{betabeamhigh,betaCERNupgrade,bblhc,CDFL07}. A number
of issues (e.g., the achievable intensities or the size of the decay
ring) need to be studied in detail in order to understand the
feasibility of these higher-$\gamma$ beta-beams and their physics
reach. Another option which emerged in the recent past is the
possibility of high-Q value beta-beams~\cite{rubbia}. These beams
exploit the decay of high Q-valued ions, namely $^{8}$B and
$^{8}$Li. The same neutrino energies can be achieved with a boost
factor 4 times smaller than for $^{18}$Ne and $^{6}$He. In order to
obtain useful luminosities at the detector, a much higher number of
beta decays is therefore required.

The intensity of the beam plays a crucial role in the physics reach of
the setup as it controls the statistics available. It depends mainly
on the production rate of the isotopes and on space-charge
limitations. At present, three possibilities for ion production have
been considered: ISOL method at medium energy and direct production
with and without a storage ring. The ISOL technique~\cite{cern-study}
uses typically (0.1--2)~GeV protons from the proposed 2.2 GeV Super
Proton Linac, which will be used to activate the nuclear reactions
producing the nuclide of interest. For $^{6}$He production, a heavy
metal target, such as mercury or water-cooled tungsten, will be used to
transform the proton beam into a neutron flux which then impacts on a 
cylinder of BeO surrounding the target. $^{6}$He is then produced via
the $^{9}$Be(n,$\alpha$) reaction. The $^{18}$Ne can be created
directly by proton spallation on a MgO target. The present studies
indicate that with a 200~kW power, one could achieve a number of ions
per second $>10^{13}/$s for $^{6}$He and $< 8 \times 10^{11}$ for
$^{18}$Ne. For direct production, low energy and high intensity ion
beams are used on solid or gas targets. Compound nuclei form at low
energy due to the high cross section and the required ions are
generated. Various preliminary studies show that a production rate of
$10^{13}$ ions per second could be achieved for $^{18}$Ne and
$^{6}$He. Direct production can be enhanced using a storage ring in
which primary ions which did not interact at the first passage are
recirculated and reaccelerated. This technique is possible thanks to
ionisation cooling~\cite{rubbia} and might allow high production rates
of $^{8}$B and $^{8}$Li, up to $10^{14}/$s for $^{8}$Li and
$10^{13}/$s for $^{8}$B.


\section{Resolving neutrino oscillation degeneracies in a neutrino
  run}

In our analysis we consider a beam of neutrinos only, but exploit the
rich oscillatory pattern of the signal. In particular, requiring a low
energy threshold for the detector, we can access more than one
oscillation maximum. This allows one to obtain a very good physics
reach and to efficiently resolve the problem of degeneracies.

The oscillation probability $P(\nu_e (\overline{\nu}_e) \rightarrow
\nu_\mu (\overline{\nu}_\mu)) \equiv P^\pm_{e \mu}$ for $\nu_e \,
(\overline{\nu}_e)$ into $\nu_\mu \, (\overline{\nu}_\mu)$ conversion
can be expanded in the small parameters $\bar \theta_{13}$,
$\Delta_{12}/\Delta_{13}$, $\Delta_{12}/A$ and
$\Delta_{12}L$~\cite{aC00}, where the shorthand $\Delta_{ji} \equiv
\Delta m_{ji}^{2}/(2E)$ is being used,
\begin{eqnarray}
\label{prob}
& P^\pm_{e \mu} (\bar \theta_{13}, \bar \delta)  =  
\sin^2 2 \bar\theta_{13} \, \sin^2 \bar \theta_{23} \left
(\frac{\Delta_{31} }{B_\mp} \right)^2\sin^2 \left (\frac{B_\mp L}{2}
\right)  + \cos^2 \bar \theta_{23} \, \sin^2 2 \bar \theta_{12} \left(
\frac{\Delta_{21}}{ A } \right )^2 \sin^2 \left(\frac{A L }{ 2 }
\right) \nonumber \\[1ex]   
& + \cos \bar \theta_{13} \, \sin 2 \bar \theta_{13} \, \sin 2
\bar \theta_{12} \, \sin 2 \bar \theta_{23} \, \frac{\Delta_{21}}{A} 
\frac{\Delta_{31} }{ B_\mp } \sin \left (\frac{A L }{ 2 } \right )\sin
\left( \frac{ B_\mp L }{ 2 } \right ) \cos\left( \pm \bar \delta -
\frac{\Delta_{31} L }{2}\right) ~, 
\end{eqnarray}
where the $\pm$ corresponds to neutrinos/anti-neutrinos and $B_\mp
\equiv A \mp\Delta_{31}$. Here we are using $A =
\sqrt{2}G_{F}\bar{n}_{e}(L)$ (the constant density approximation for
the index of refraction) where $\bar{n}_{e} =
1/L\int_{0}^{L}n_{e}(L')dL'$ is the average electron density and
$n_{e}(L)$ is the electron density along the baseline. 

The number of neutrino (antineutrino) events in the $i$-th neutrino
(antineutrino) energy bin for a given pair (${\bar \theta}_{13},{\bar
  \delta}$) is given by
\begin{equation}
\label{E:nneutr}
N_i (\bar \theta_{13},\bar \delta) =  \mathcal{N}_{\rm T} \, t \,
  \int_{E_i}^{E_i + \Delta E} \, \epsilon (E_\nu) \, \,
  \sigma_{\nu_\mu(\overline{\nu}_\mu)} (E_\nu) \, \, 
  P^{\pm}_{e\mu} (E_\nu,\bar \theta_{13},\bar \delta) \, \, 
  \Phi_{\nu_e (\overline{\nu}_e)} (E_\nu) \, \, dE_\nu ~,
\end{equation}
where $\mathcal{N}_{\rm T}$ is the number of targets in the detector,
$t$ is the time of data taking, $\epsilon(E_\nu)$ is the detector
efficiency, $\sigma (E_\nu)$ is the interaction cross section, $\Phi
(E_\nu)$ is the beam spectrum and $\Delta E$ is the neutrino energy
resolution of the detector. As it is apparent from Eq.~(\ref{prob}),
the extraction of $\theta_{13}$, the sign of \deltaatm, $\delta$ and
$\theta_{23}$ from Eq.~(\ref{E:nneutr}) suffers from the problem of 
degeneracies~\cite{deg1,deg2,deg3,deg4,deg5}. In addition to the true
solution ($\bar \theta_{13}$, $\bar \delta$,
$\mathrm{sign}(\Delta m^2_{31})$, $\bar \theta_{23}$), other fake
(clone) solutions ($\theta_{13}$, $\delta$, $\pm \mathrm{sign}( \Delta
m^2_{31})$, $\theta_{23}$) are allowed.

In order to get an analytical understanding of the degeneracies, we
consider the simplified case of infinite energy resolution, for which
the integrals in Eq.~(\ref{E:nneutr}) reduce to products. We compare
the number of events at the neutrino energy of the first oscillation
maximum, $E_1$, with the number at the second oscillation maximum,
$E_2$. In the approximation considered, we have $N_1 (E_1) = c_1
P_{e\mu}^+ (E_1)$ and $N_2 (E_2) = c_2 P_{e\mu}^+ (E_2)$, where
$c_1$ and $c_2$ are the product of the number of targets, time of data
taking, efficiency, the neutrino flux and cross section at each of the
two energies, respectively. Although a more realistic analysis should
be performed taking into account the finite energy resolution,
statistical and systematic errors, backgrounds and the full
differential cross section for the detection processes, we use this 
simplified approach to show the potentialities of using the spectral
information with only one polarity. As it will be shown in Section V,
a more detailed analysis confirms the qualitative results obtained
here.

Let us first consider the intrinsic degeneracy, for which given the
true value of the pair $(\bar \theta_{13}, \bar \delta)$, the clone
solution $(\theta_{13}, \delta)$ is located by solving~\cite{deg1}
\begin{eqnarray}
N_1 (\bar \theta_{13}, \bar \delta, \mathrm{sign}( \Delta
m^2_{31}), \bar \theta_{23}) 
& = & N_1 (\theta_{13},\delta,\mathrm{sign}( \Delta m^2_{31}),
\bar \theta_{23}) ~,\\[1ex]  
N_2 (\bar \theta_{13}, \bar \delta,  \mathrm{sign}( \Delta
m^2_{31}), \bar \theta_{23})
& = & N_2 (\theta_{13},\delta,\mathrm{sign}( \Delta m^2_{31}),
\bar \theta_{23})   ~. 
\end{eqnarray}
As it is straightforward to show, only one solution is allowed for
$\theta_{13}$ and $\delta$ for a measured number of events $N_1$ and
$N_2$. The CP-violating effects become more important at low energy
and in particular at the second oscillation maximum. Therefore, the
intrinsic degeneracy is fully resolved by exploiting the neutrino
signal at first and second oscillation maximum. In the case of a
neutrino and antineutrino beam considered at first oscillation
maximum, a similar result is obtained.

We study next the sign degeneracy. The clone solution
satisfies~\cite{deg2}
\begin{eqnarray}
 N_1 (\bar \theta_{13}, \bar \delta, \mathrm{sign}(
\Delta m^2_{31}), \bar \theta_{23})
& = & N_1 (\theta_{13}, \delta, -\mathrm{sign}( \Delta m^2_{31}),
\bar \theta_{23}) ~, \\[1ex] 
N_2 (\bar \theta_{13}, \bar \delta, \mathrm{sign}(
\Delta m^2_{31}), \bar \theta_{23})
& = & N_2 (\theta_{13}, \delta, -\mathrm{sign}( \Delta m^2_{31}),
\bar \theta_{23}) ~, 
\end{eqnarray}
and is found to be
\begin{eqnarray}
\sin^2 2 \theta_{13} & \simeq & \sin^2  2 \bar \theta_{13} \left( 1
+ 4 \frac{A}{\Delta_{31}} \right) ~,\\[1ex] 
\sin  \delta & \simeq & \sin \bar \delta~,
\end{eqnarray}
where we have only kept terms up to first order in
$A/\Delta_{31}$. This shows that the sign degeneracy affects only
mildly the determination of $\bar \theta_{13}$ and very weakly that of
$\bar \delta$. The signal at first and second oscillation maximum is
not sufficient to determine the type of neutrino mass
ordering. However, including information on the neutrino oscillation
probability at other energies breaks this degeneracy. In particular,
matter effects increase with energy and the high energy bins turn out
to play a very important role in breaking the sign degeneracy. In the
case of combining the first oscillation maximum for neutrinos and
antineutrinos, we have also a clone solution, with information from
additional energy bins also necessary to fully resolve this degeneracy.

Finally, we study the octant degeneracy, for which the clone
solution is such that~\cite{deg3}
\begin{eqnarray}
N_1 (\bar \theta_{13}, \bar \delta,  \mathrm{sign}( \Delta
m^2_{31}),\bar \theta_{23})
& = & N_1 (\theta_{13}, \delta,  \mathrm{sign}( \Delta m^2_{31}),
\pi/2 - \bar \theta_{23}) ~, \\[1ex] 
N_2 (\bar \theta_{13}, \bar \delta,  \mathrm{sign}( \Delta
m^2_{31}), \bar \theta_{23})
& = & N_2 (\theta_{13}, \delta,  \mathrm{sign}( \Delta m^2_{31}),
\pi/2 - \bar \theta_{23})~, 
\end{eqnarray}
and is given by 
\begin{eqnarray}
\sin^2 2 \theta_{13} & \simeq & \tan^2 \bar \theta_{23} \,
\sin^2 2 \bar \theta_{13} + (1- \tan^2 \bar \theta_{23}) 
\sin^2 2 \bar \theta_{12} \left(
\frac{\Delta_{21}}{\Delta_{31}}\right)^2 \frac{\pi^2}{4} ~, \\[1ex] 
\sin \delta & \simeq & \frac{\sin \bar \delta}{\tan \bar \theta_{23}}
\, \left( 1 + \frac{\pi^2}{8} \, \left(1 - \frac{1}{\tan^2 \bar
  \theta_{23}}\right) \, \frac{\sin^2 2 \bar \theta_{12}}{\sin^2 2 \bar
  \theta_{13}} \, \left(\frac{\Delta_{21}}{\Delta_{31}}\right)^2
\right) ~.
\end{eqnarray}
This is valid in the whole allowed range of the oscillation parameters
for $\sin^2 2 \bar \theta_{13} > 10^{-3}$ and only terms up to
$\mathcal{O}(\Delta_{21}/\Delta_{31})^2$ have been retained. As it has
been discussed~\cite{lowNuFact}, the information from the low energy
bins plays a crucial role in resolving this degeneracy. Similar
considerations can be done for the case of a neutrino and antineutrino
run.

Our simplified analysis suggest that even with the neutrino run alone,
by exploiting the oscillatory pattern of the signal, it is possible to
resolve degeneracies and obtain a very good sensitivity to the unknown
neutrino parameters. In the following, we substantiate these claims
with a detailed numerical analysis.


\section{Numerical simulations and analysis of the data}

\begin{figure}[t]
\begin{center}
\includegraphics[width=3.8in]{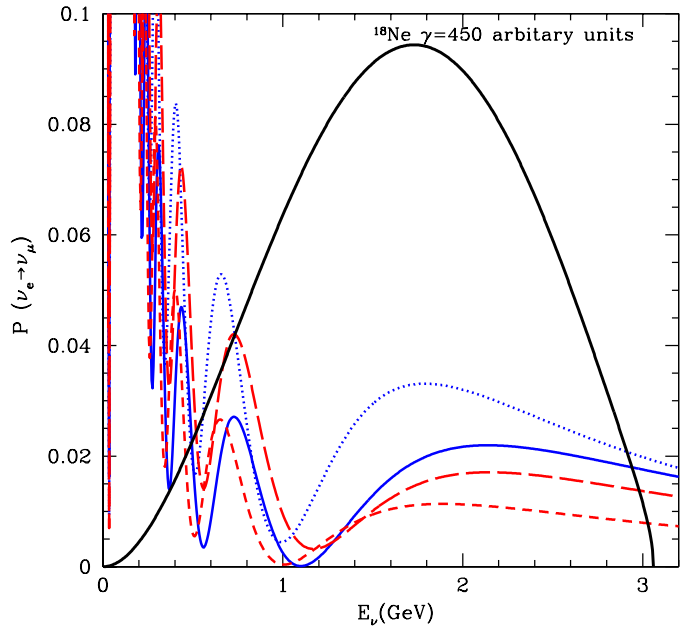}
\caption{\textit{Transition probability of $\nu_e$ into $\nu_\mu$ as a
function of the neutrino energy for normal hierarchy and
$\delta=0^{\circ}$ (blue solid line), normal hierarchy and
$\delta=90^{\circ}$ (blue dotted line), inverted hierarchy and
$\delta=0^{\circ}$ (red short dashed line) and inverted hierarchy and
$\delta=90^{\circ}$ (red long dashed line). Also shown in arbitrary
units the unoscillated beta-beam neutrino spectrum from $^{18}$Ne
decays and $\gamma=450$.}}
\label{fig:spectrumprob}
\end{center}
\end{figure}

The physics strategy followed here exploits electron neutrino beams
from boosted $^{18}$Ne $\beta^{+}$ decays for a single
baseline. Future CERN accelerator facilities could provide the 
production environment. The neutrino detector would be ideally placed
at Boulby, located at $L=1050$~km from CERN. The Lorentz factor we
assume here corresponds to a conservative (for the upgraded SPS)
$\gamma = 450$, for which the mean electron neutrino energy is $\langle
E_\nu \rangle \simeq \gamma E_{0}\sim 1.5$~GeV, $E_{0} = 3.41$~MeV
being the positron end-point energy for $^{18}$Ne. With such a setup,
both the first and second oscillation maxima of the appearance
probability could, in principle, be studied. This is illustrated in
Fig.~\ref{fig:spectrumprob} where we show the transition probability
of $\nu_e$ into $\nu_\mu$ for two values of the CP-violating phase
$\delta$ and for both mass orderings.

The results of this study are shown for two possible experimental
scenarios, which only differ in their statistics. This analysis allows
us to quantify the benefits of increased detector sizes, ion
intensities and/or exposure times. We first consider an exposure 
corresponding to $10^{21}$ ions-kton-years. This could be obtained,
for example, assuming $2 \times 10^{18}$ useful ion decays per year
and a 50~kton detector located at Boulby with 10~years of data
taking. As we will show, the statistics plays a crucial role in the
physics reach of the setup. Therefore, we also study a rather
optimistic scenario, obtained by upgrading the first scenario by a 
factor of five in statistics, i.e., with an exposure of $5 \times
10^{21}$ ions-kton-years. Considering that the time of data taking
cannot be substantially extended, this exposure could be achieved by
using a larger detector or increasing the ion luminosity or
both. Obviously, the results for exposuree.s larger than the first
scenario but not as quite optimistic as the second one will
interpolate between the two. Possible detector technologies considered
in the literature include water \v{C}erenkov, liquid argon, totally
active scintillator or iron calorimeter detectors. For our purposes
low energy threshold and good neutrino energy resolution are
requirements for the choice of detector technology. For instance,
liquid argon and totally active scintillator detectors might have
these characteristics.

As mentioned above, a crucial aspect of the analysis performed here is
to use the spectral information, and hence the energy binning of the
signal becomes fundamental. As carefully described in the previous
section, our main strategy consists in exploiting simultaneously many
$E/L$'s, and therefore we assume a $200$~MeV bin width and an energy
detection threshold of $400$~MeV. In what follows, the muon-neutrino
appearance signal is binned in eleven bins with a bin width of
$200$~MeV in the [0.4, 2.0] GeV energy range, plus a unique, last bin,
filled with the neutrino events from $2.0$~GeV up to the end point of
the spectrum at $3.06$~GeV. For our numerical analysis, we use the
following $\chi^{2}$ definition:

\begin{equation}
\chi^2 = \sum_{i,j} \; (n_{i} - N_{i}) C_{ij}^{-1} (n_{j} - N_{j})\,,
\label{eq:chi}
\end{equation}
\noindent
where $N_{i}$ is the predicted number of muons for a certain
oscillation hypothesis, $n_{i,p}$ are the simulated ``data'' from a 
Gaussian or Poisson smearing. The $2 N_{\rm bin} \times 2 N_{\rm bin}$
covariance matrix $C$, which is given by 

\begin{equation}
C_{i,j}^{-1}\equiv \delta_{ij}(\delta n_{i})^2 
\end{equation}
\noindent
where $(\delta n_{i}) = \sqrt{n_{i} + (f_{sys}\cdot n_{i})^2}$,
contains both statistical and a $2\%$ overall systematic error
($f_{sys}=0.02$). The confidence level (CL) contour plots presented in
the figures in the next section have been calculated for 2 degrees of
freedom (d.o.f.) statistics. 

\begin{figure}[t]
\begin{center}
\begin{tabular}{ll}
\includegraphics[width=3.2in]{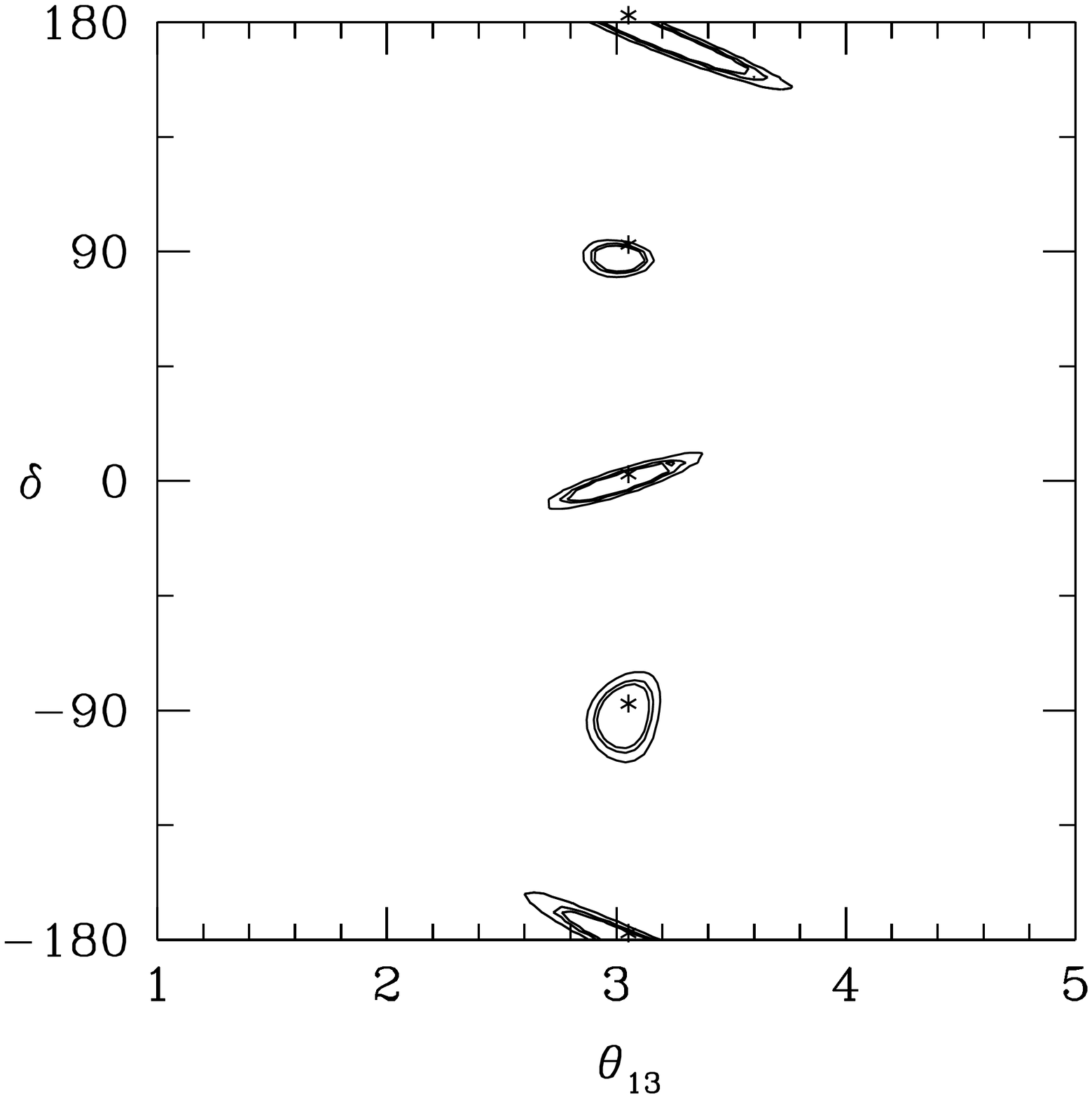}&\hskip 0.cm
\includegraphics[width=3.2in]{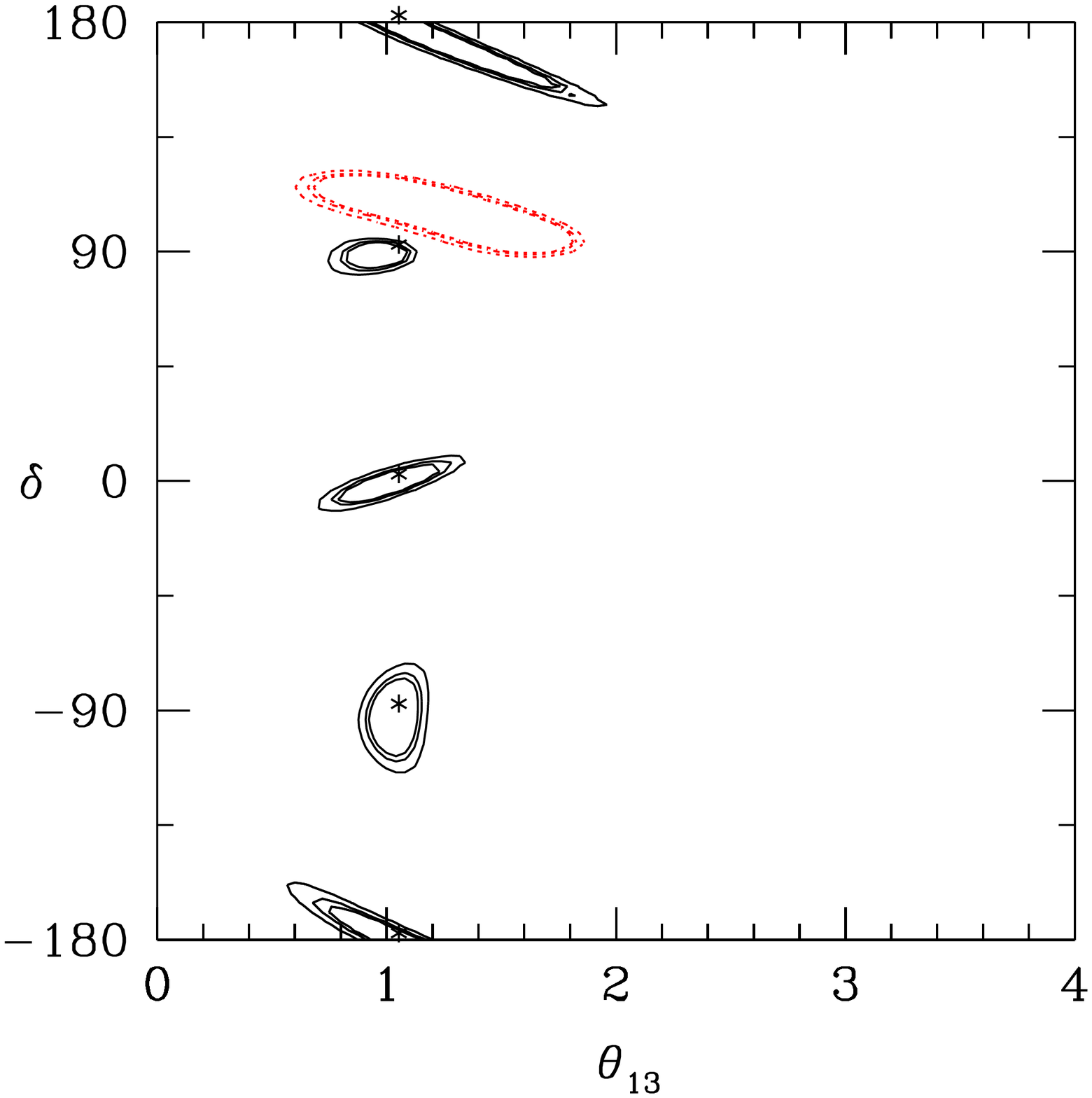}\\
\end{tabular}
\caption[]{\textit{For an exposure of $5 \times
    10^{21}$~ions-kton-years, $90\%$, $95\%$ and $99\%$ (for 2 d.o.f)
    CL contours resulting from the fits if the true values Nature has
    chosen are $\theta_{13}=3^\circ$ (left panel) or
    $\theta_{13}=1^\circ$ (right panel), and $\delta=0^{\circ}$,
    $90^{\circ}$, $-90^{\circ}$ or $180^{\circ}$. Dashed-red contours
    represent the hierarchy-clone solution.}}
\label{fig:deg}
\end{center}
\end{figure}

Realistic background assumptions have been included when computing the
simulated data. We have considered two types of backgrounds: an
intrinsic beam-induced background (taken as a constant fraction,
$0.1\%$, of the unoscillated events) plus the atmospheric neutrino
contribution. In the energy range of interest, there are about $30$
atmospheric neutrino interactions per kton-year which could mimic a
muon coming from the oscillated $\nu_e \to
\nu_\mu$~\cite{mauro2,MINOSatm,andy}. Assuming a beam duty factor of
$10^{-3}$, the number of muon background events induced by atmospheric 
muon neutrino interactions  would be $\sim 0.03$ per kton-year, to be
rescaled accordingly to the detector size and the exposure time. We
think the treatment of the backgrounds presented here is
conservative.


\section{Results}

\begin{figure}[t]
\begin{center}
\includegraphics[width=3.8in]{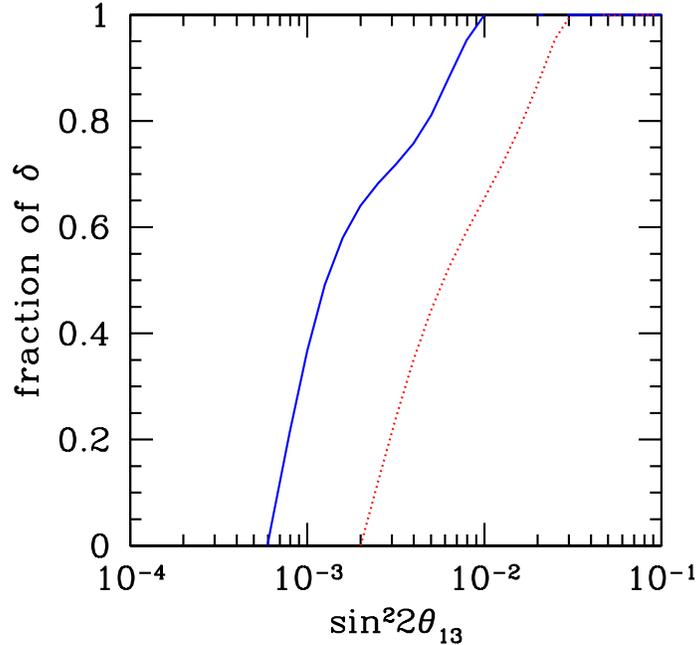}

\caption[]{\textit{$99\%$ CL hierarchy resolution (2 d.o.f). The
    dotted red curve depicts the results assuming an exposure of
    $10^{21}$~ion-kton-years in Boulby.  The solid blue curve depicts
    the results assuming the statistics quoted before is improved by a
    factor of five.}}
\label{fig:hier}
\end{center}
\end{figure}

\begin{figure}[t]
\begin{center}
\includegraphics[width=4.in]{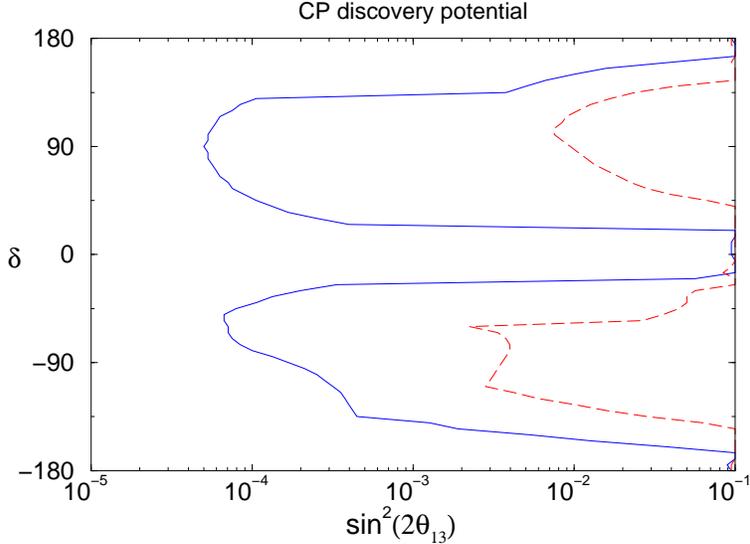}
\caption[]{\textit{Discovery at $99\%$ CL of CP violation (2
    d.o.f). The dashed red curve depicts the results assuming an
    exposure of $10^{21}$~ions-kton-years in Boulby. The solid blue
    curve depicts the results for an exposure of $5 \times
    10^{21}$~ions-kton-years.}}  
\label{fig:cp}
\end{center}
\end{figure}

We present in Fig.~\ref{fig:deg} the $90\%$, $95\%$ and $99\%$ CL
contours for a fit to the simulated data from the beta-beam experiment
described in the previous section. The ``true'' parameter values that
we have chosen for these examples are depicted in the figures with a
star: we have explored four different values of $\delta=0^{\circ}$,
$90^{\circ}$, $-90^{\circ}$ and $180^{\circ}$ and two possible values
of $\tetaot=3^\circ$ (left panel) and $1^\circ$ (right panel). The
simulations are for the normal mass hierarchy and $\theta_{23}$ in the
first octant ($\sin^2 \theta_{23} = 0.41$ which corresponds to
$\theta_{23}=40^\circ$). The statistics considered corresponds to the
optimistic high-statistics scenario, with an exposure of $5 \times
10^{21}$ ions-kton-years. The analysis depicted in Fig.~\ref{fig:deg}
includes the study of the discrete degeneracies. That is, we have
fitted the data assuming the wrong hierarchy (i.e., negative hierarchy)
and the additional clone solutions (if present) are shown in dashed
red. We have also considered the impact of the wrong choice for the
$\theta_{23}$ octant (we fitted the data assuming $\sin^2
\theta_{23}=0.59$, which corresponds to $\theta_{23}=50^\circ$). 
Notice that in Fig.~\ref{fig:deg} the $\theta_{23}$-octant ambiguity
is solved at the $99\%$ CL for the values of $\delta$ illustrated. The
additional solutions associated to the wrong choice of the mass
hierarchy are not present at the same CL if $\tetaot$ is small enough,
i.e., $\theta_{13}<6^\circ$. For \emph{larger} values of $\theta_{13}$
the sign degeneracy is present for some values of the CP-violating
phase $\delta$, but its location is very similar to the simulated true
value and therefore the presence of these degeneracies will hardly
interfere with the measurement of CP violation. This behaviour of the
$\theta_{23}$ degenerate solutions is opposite to the normal case, in
which the resolution of the degeneracies gets harder as the value of
$\theta_{13}$ decreases. The reason for that is the enormous impact of
the solar term at the lower energies and the long baseline exploited
here. For instance, for $\langle E_\nu \rangle \simeq 1$~GeV, the
solar term contribution is larger than that of the atmospheric term
for all the values of $\theta_{13}<6^\circ$.

\begin{figure}[t]
\begin{center}
\includegraphics[width=4.in]{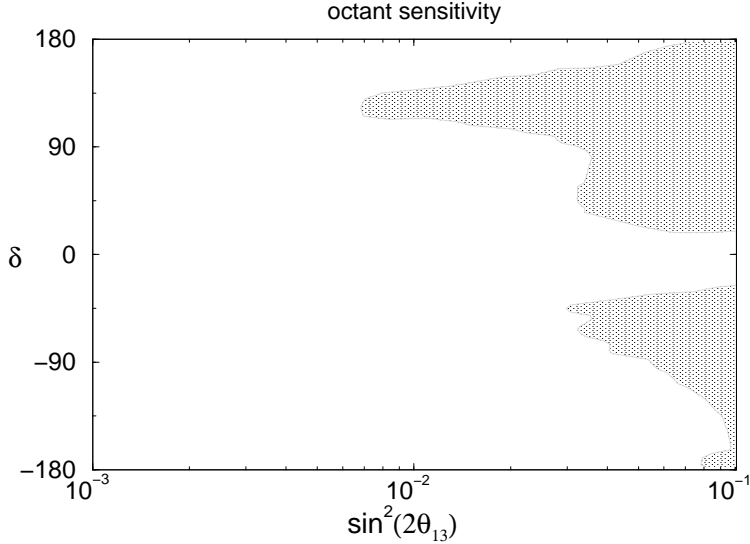}
\caption{\textit{$99\%$ CL determination of the $\theta_{23}$ octant.
The grey region denotes the parameter space for which the octant
degeneracy is not resolved at the $99\%$ CL. Results are shown for an
exposure of $5 \times 10^{21}$~ions-kton-years.}}
\label{fig:teta}
\end{center}
\end{figure}

Figs.~\ref{fig:hier}, \ref{fig:cp} and \ref{fig:teta} summarise, for
the low- and high-statistics scenarios, the physics reach of the
beta-beam experiment considered here. The analysis takes into
account the impact of both the intrinsic and discrete
degeneracies. Fig.~\ref{fig:hier} shows the region in the ($\sin^2 2  
\theta_{13}$, ``fraction of $\delta$'') plane for which the mass
hierarchy can be resolved at the $99\%$ CL (2 d.o.f). Note that, with
a background level of $0.1\%$ and a beam duty cycle of $10^{-3}$, the
hierarchy can still be determined in both scenarios if $\sin^2 2
\theta_{13}>0.03$ (i.e., $\theta_{13}> 5^\circ$) for all values of the
CP-violating phase $\delta$. Fig.~\ref{fig:cp} shows the region in the
($\sin^2 2 \theta_{13}$, $\delta$) plane for which a given (non-zero)
value of the CP-violating phase $\delta$ can be distinguished at
the $99\%$ CL (2 d.o.f.) from the CP-conserving case, i.e., $\delta =0,
\pm 180^\circ$. The results are given for both the low- and
high-statistics scenarios. Again, even in the presence of non
negligible beam-induced plus atmospheric background levels, the
CP-violating phase $\delta$ could be measured with a $99\%$ CL error
smaller than $\sim 20^\circ$ if $\sin^2 2 \theta_{13}>10^{-3}$ in the
high-statistics scenario. Finally, the white area in
Fig.~\ref{fig:teta} represents the region in the ($\sin^2 2
\theta_{13}$, $\delta$) plane for which the octant in which
$\theta_{23}$ lies can be determined at the $99\%$ CL (2 d.o.f). The
result is illustrated for the high-statistics scenario. In general,
the resolution of the $\theta_{23}$ octant ambiguity is extremely
difficult, and in order to eliminate this degeneracy, combining data
from different experiments might be crucial~\cite{davide}. Here, we
benefit from the solar term contribution in the lower energy bins and
therefore, the octant ambiguity is resolved for relatively small
values of $\theta_{13}$, if the statistics is high enough. For the low
luminosity scenario, the degeneracy is harder to resolve and it is
present (at the $99\%$ CL) in almost all the parameter space explored
here.


\section{Summary and Conclusions}

In the present article, we have studied the physics reach of a
beta-beam with intermediate $\gamma$ and long baseline. We have
considered a neutrino beam sourced by $^{18}$Ne decays with $\gamma =
450$ and a baseline of 1050~km corresponding to the CERN-Boulby mine
distance. This choice of setup is motivated by the recent studies
about the possible upgrade of the SPS at CERN which would allow to
obtain high boost factors for the ions and by the possibility of
locating few-ten-kton size detectors at the Boulby mine.

In a qualitative way, we have analytically studied the capability that
a neutrino run alone could have to resolve the problem of degeneracies
and we have shown that, by exploiting the oscillatory behaviour of the
signal, it is possible to fully resolve such degeneracies in a large
part of the allowed parameter space. We have performed a numerical
analysis simulating the data for a future beta-beam experiment with an
exposure of $10^{21}$ useful ion decays-kton-years. Realistic
backgrounds (intrinsic plus atmospheric neutrino contamination) and
beam-duty factor ($10^{-3}$) have been considered when computing the
$\chi^2$ function used for this analysis. By exploiting the neutrino
data only, the beta-beam setup presented here can determine the type
of neutrino mass hierarchy at the $99\%$~CL, regardless of the value
of the CP-violating phase, for values of the mixing parameter $\sin^2
2 \theta_{13}$ larger than $0.03$, and establish leptonic CP violation
if $\sin^2 2\theta_{13} > 0.01$.

If the mixing angle $\theta_{13}$ turns out to be very small, a quest
for physics answers would evidently require an increase of the
exposure quoted above. We illustrate the physics reach of this
beta-beam setup assuming a factor of five improvement in the
statistics, considering $5 \times 10^{21}$ ions-kton-years. In this
high-luminosity scenario, the value of the CP-violating phase $\delta$
can be measured with a $99\%$~CL error of $\sim 20^\circ$ if $\sin^2 2
\theta_{13} > 10^{-3}$, with some sensitivity down to very small
values of $\sin^2 2 \theta_{13} \simeq 10^{-4}$ . If $\theta_{13}$
turns out to be small, the octant degeneracy can be resolved through
the solar mixing term contribution.
 
In summary, we have analysed a beta-beam setup with a neutrino run and
we have found that, by exploiting the oscillatory behaviour of the
signal, this option provides a very good overall physics reach,
competitive with other setups explored in the literature.

\section*{Acknowledgements} 
We thank Adrian Fabich and Mats Lindroos for useful discussions since
the initial stages of this work. OM thanks the IPPP for hospitality
during the completion of this study. OM was supported by the European
Programme ``The Quest for Unification'' under contract
MRTN-CT-2004-503369. CO acknowledges the support of a STFC
studentship. SPR is partially supported by the Spanish Grant
FPA2005-01678 of the MCT. SP acknowledges the support of CARE,
contract number RII3-CT-2003-506395.


\begin{thebibliography}{99}

\bibitem{atm}
  Y.~Fukuda {\it et al.}  [Super-Kamiokande Collaboration],
  Phys.\ Rev.\ Lett.\  {\bf 81} (1998) 1562 
  [arXiv:hep-ex/9807003];
%
  M.~Ambrosio {\it et al.}  [MACRO Collaboration],
  Phys.\ Lett.\  B {\bf 434} (1998) 451 
  [arXiv:hep-ex/9807005];
%
  M.~C.~Sanchez {\it et al.}  [Soudan 2 Collaboration],
  Phys.\ Rev.\  D {\bf 68} (2003) 113004 
  [arXiv:hep-ex/0307069].


\bibitem{SKatm}
  Y.~Ashie {\it et al.}  [Super-Kamiokande Collaboration],
  Phys.\ Rev.\ D {\bf 71} (2005) 112005
  [arXiv:hep-ex/0501064].


\bibitem{sol}
  B.~T.~Cleveland {\it et al.},
  Astrophys.\ J.\  {\bf 496} (1998) 505;
%
  Y.~Fukuda {\em et al.} [Kamiokande Collaboration],
  Phys.\ Rev.\ Lett.\  {\bf 77} (1996) 1683;
%
  J.~N.~Abdurashitov {\it et al.}  [SAGE Collaboration],
  J.\ Exp.\ Theor.\ Phys.\  {\bf 95} (2002) 181 
  [Zh.\ Eksp.\ Teor.\ Fiz.\  {\bf 122} (2002) 211]
  [arXiv:astro-ph/0204245];
%
  W.~Hampel {\it et al.}  [GALLEX Collaboration],
  Phys.\ Lett.\ B {\bf 447} (1999) 127;
%
  T.~A.~Kirsten  [GNO Collaboration],
  Nucl.\ Phys.\ Proc.\ Suppl.\  {\bf 118} (2003) 33.


\bibitem{SKsolar}
  S.~Fukuda {\it et al.}  [Super-Kamiokande Collaboration],
  Phys.\ Lett.\ B {\bf 539} (2002) 179
  [arXiv:hep-ex/0205075].


\bibitem{SNO}
  Q.~R.~Ahmad {\it et al.}  [SNO Collaboration],
  Phys.\ Rev.\ Lett.\  {\bf 87} (2001) 071301 
  [arXiv:nucl-ex/0106015];
%
  {\it ibid.} {\bf 89} (2002) 011301 
  [arXiv:nucl-ex/0204008];
%
  and {\it ibid.} {\bf 89} (2002) 011302 
  [arXiv:nucl-ex/0204009];
%
  S.~N.~Ahmed {\it et al.} [SNO Collaboration],
  Phys.\ Rev.\ Lett.\ {\bf 92} (2004) 181301 
  [arXiv:nucl-ex/0309004];
%
  B.~Aharmim {\it et al.}  [SNO Collaboration],
  Phys.\ Rev.\ C {\bf 72} (2005) 055502 
  [arXiv:nucl-ex/0502021].


\bibitem{CHOOZ}
  M.~Apollonio {\it et al.}  [CHOOZ Collaboration],
  Phys.\ Lett.\  B {\bf 466} (1999) 415 
  [arXiv:hep-ex/9907037];
%
  M.~Apollonio {\it et al.}  [CHOOZ Collaboration],
  Eur.\ Phys.\ J.\  C {\bf 27} (2003) 331 
  [arXiv:hep-ex/0301017].


\bibitem{PaloVerde}
  F.~Boehm {\it et al.},
  Phys.\ Rev.\ Lett.\  {\bf 84} (2000) 3764 
  [arXiv:hep-ex/9912050];
%
  Phys.\ Rev.\  D {\bf 62} (2000) 072002 
  [arXiv:hep-ex/0003022];
%
  and {\it ibid.} {\bf 64} (2001) 112001 
  [arXiv:hep-ex/0107009].


\bibitem{KamLAND}
  K.~Eguchi {\it et al.}  [KamLAND Collaboration],
  Phys.\ Rev.\ Lett.\  {\bf 90} (2003) 021802 
  [arXiv:hep-ex/0212021];
%
  T.~Araki {\it et al.}  [KamLAND Collaboration],
  Phys.\ Rev.\ Lett.\  {\bf 94} (2005) 081801 
  [arXiv:hep-ex/0406035].


\bibitem{K2K}
  M.~H.~Ahn {\it et al.}  [K2K Collaboration],
  Phys.\ Rev.\  D {\bf 74} (2006) 072003 
  [arXiv:hep-ex/0606032].


\bibitem{MINOS}
  D.~G.~Michael {\it et al.}  [MINOS Collaboration],
  Phys.\ Rev.\ Lett.\  {\bf 97} (2006) 191801 
  [arXiv:hep-ex/0607088].


\bibitem{TSchwNuFact07}
  M.~Maltoni, T.~Schwetz, M.~A.~Tortola and J.~W.~F.~Valle,
  New J.\ Phys.\  {\bf 6} (2004) 122 
  [arXiv:hep-ph/0405172v6];
%
  T.~Schwetz,
  arXiv:0710.5027 [hep-ph].


\bibitem{ISS}
  A.~Bandyopadhyay {\it et al.}  [ISS Physics Working Group],
  arXiv:0710.4947 [hep-ph].


\bibitem{Euronu}
  ``Euro-$\nu$: High Intensity Neutrino Oscillation Facility in
  Europe'', FP7-infrastructures-2007-1 project number 212372.


\bibitem{T2K}
  Y.~Hayato {\it et al.},
  Letter of Intent, available at 
\url{http://neutrino.kek.jp/jhfnu/}


\bibitem{newNOvA}
  D.~S.~Ayres {\it et al.}  [NOvA Collaboration],
  hep-ex/0503053.
  FERMILAB-PROPOSAL-0929, March 21, 2005.
  Revised \nova Proposal available at
\url{http://www-nova.fnal.gov/NOvA_Proposal/Revised_NOvA_Proposal.html}


\bibitem{Barger:2007jq}
  V.~Barger, P.~Huber, D.~Marfatia and W.~Winter,
  Phys.\ Rev.\  D {\bf 76} (2007) 053005 
  [arXiv:hep-ph/0703029];
%
  V.~Barger {\it et al.},
  arXiv:0705.4396 [hep-ph].

  
\bibitem{nufact}
  S.~Geer,
  Phys.\ Rev.\  D {\bf 57} (1998) 6989 
  [Erratum-ibid.\  D {\bf 59} (1999) 039903]
  [arXiv:hep-ph/9712290];
%
  A.~De R\'ujula, M.~B.~Gavela and P.~Hern\'andez,
  Nucl.\ Phys.\ B {\bf 547} (1999) 21 
  [arXiv:hep-ph/9811390];
%
  V.~Barger, S.~Geer, R.~Raja and K.~Whisnant,
  Phys.\ Rev.\ D {\bf 62} (2000) 013004 
  [arXiv:hep-ph/9911524];
%
  M.~Freund, M.~Lindner, S.~T.~Petcov and A.~Romanino, 
  Nucl.\ Phys.\ B {\bf 578} (2000) 27 
  [arXiv:hep-ph/9912457].


 \bibitem{zucchelli}
  P.~Zucchelli,
  Phys.\ Lett.\ B {\bf 532} (2002) 166.


\bibitem{mauro1}
  M.~Mezzetto,
  J.\ Phys.\ G {\bf 29} (2003) 1781 
  [arXiv:hep-ex/0302005].


\bibitem{mauro2}
  M.~Mezzetto,
  J.\ Phys.\ G {\bf 29} (2003) 1771 
  [arXiv:hep-ex/0302007].


\bibitem{deg1}
  J.~Burguet-Castell {\it et al.}, 
  Nucl.\ Phys.\  B {\bf 608} (2001) 301
  [arXiv:hep-ph/0103258].


\bibitem{deg2}
  H.~Minakata and H.~Nunokawa,
  JHEP {\bf 0110} (2001) 001 
  [arXiv:hep-ph/0108085].


\bibitem{deg3} 
  G.~L.~Fogli and E.~Lisi,
  Phys.\ Rev.\ D {\bf 54} (1996) 3667 
  [arXiv:hep-ph/9604415].


\bibitem{deg4} 
  V.~Barger, D.~Marfatia and K.~Whisnant,
  Phys.\ Rev.\  D {\bf 65} (2002) 073023 
  [arXiv:hep-ph/0112119].


\bibitem{deg5}
  M.~Freund, P.~Huber and M.~Lindner,
  Nucl.\ Phys.\ B {\bf 615} (2001) 331 
  [arXiv:hep-ph/0105071];
%
  T.~Kajita, H.~Minakata and H.~Nunokawa,
  Phys.\ Lett.\ B {\bf 528} (2002) 245
  [arXiv:hep-ph/0112345];
%
  P.~Huber, M.~Lindner and W.~Winter,
  Nucl.\ Phys.\ B {\bf 645} (2002) 3 
  [arXiv:hep-ph/0204352];
%
  H.~Minakata, H.~Nunokawa and S.~J.~Parke,
  Phys.\ Rev.\ D {\bf 66} (2002) 093012 
  [arXiv:hep-ph/0208163];
%
  A.~Donini, D.~Meloni and S.~Rigolin,
  JHEP {\bf 0406} (2004) 011 
  [arXiv:hep-ph/0312072];
%
  M.~Aoki, K.~Hagiwara and N.~Okamura,
  Phys.\ Lett.\ B {\bf 606} (2005) 371 
  [arXiv:hep-ph/0311324];
%
  O.~Yasuda,
  New J.\ Phys.\  {\bf 6} (2004) 83 
  [arXiv:hep-ph/0405005];
%
  O.~Mena and S.~J.~Parke,
  Phys.\ Rev.\ D {\bf 72} (2005) 053003 
  [arXiv:hep-ph/0505202].


\bibitem{otherexp1}
  P.~Huber, M.~Lindner and W.~Winter,
  Nucl.\ Phys.\ B {\bf 654} (2003) 3 
  [arXiv:hep-ph/0211300].


\bibitem{otherexp2}
  H.~Minakata, H.~Nunokawa and S.~J.~Parke,
  Phys.\ Rev.\ D {\bf 68} (2003) 013010 
  [arXiv:hep-ph/0301210].


\bibitem{otherexp3}
  V.~Barger, D.~Marfatia and K.~Whisnant,
  Phys.\ Lett.\ B {\bf 560} (2003) 75 
  [arXiv:hep-ph/0210428].


\bibitem{otherexp4}
  Y.~F.~Wang {\it et al.}, 
  [VLBL Study Group H2B-4],
  Phys.\ Rev.\  D {\bf 65} (2002) 073021 
  [arXiv:hep-ph/0111317];
%
  J.~Burguet-Castell {\it et al.},
  Nucl.\ Phys.\ B {\bf 646} (2002) 301 
  [arXiv:hep-ph/0207080];
%
  K.~Whisnant, J.~M.~Yang and B.~L.~Young,
  Phys.\ Rev.\ D {\bf 67} (2003) 013004 
  [arXiv:hep-ph/0208193];
%
  P.~Huber, M.~Lindner, T.~Schwetz and W.~Winter,
  Nucl.\ Phys.\ B {\bf 665} (2003) 487 
  [arXiv:hep-ph/0303232];
%
  P.~Huber {\it et al.},
  Phys.\ Rev.\ D {\bf 70} (2004) 073014 
  [arXiv:hep-ph/0403068];
%
  A.~Donini, E.~Fern\'andez-Mart\'{\i}nez and S.~Rigolin,
  Phys.\ Lett.\ B {\bf 621} (2005) 276 
  [arXiv:hep-ph/0411402];
%
  M.~Narayan and S.~Uma Sankar,
  Phys.\ Rev.\  D {\bf 61} (2000) 013003 
  [arXiv:hep-ph/9904302].


\bibitem{otherexpbeta}
  A.~Donini, 
  Nucl.\ Phys.\  B {\bf 710} (2005) 402 
  [arXiv:hep-ph/0406132];
%
  A.~Donini, E.~Fernandez-Martinez and S.~Rigolin,
  Phys.\ Lett.\  B {\bf 621} (2005) 276 
  [arXiv:hep-ph/0411402].


\bibitem{otherexp5}
  P.~Huber, M.~Maltoni and T.~Schwetz,
  Phys.\ Rev.\ D {\bf 71} (2005) 053006 
  [arXiv:hep-ph/0501037].


\bibitem{otherexp6}
  O.~Mena and S.~J.~Parke,
  Phys.\ Rev.\ D {\bf 70} (2004) 093011 
  [arXiv:hep-ph/0408070];
%
  O.~Mena,
  Mod.\ Phys.\ Lett.\ A {\bf 20} (2005) 1 
  [arXiv:hep-ph/0503097];
%
  O.~Mena, H.~Nunokawa and S.~J.~Parke,
  Phys.\ Rev.\  D {\bf 75} (2007) 033002 
  [arXiv:hep-ph/0609011];
%
  O.~Mena,
  hep-ph/0609031;
%
  A.~Jansson, O.~Mena, S.~Parke and N.~Saoulidou,
  arXiv:0711.1075 [hep-ph].


\bibitem{otherexp7}
  A.~Blondel {\it et al.},
  Acta Phys.\ Polon.\  B {\bf 37} (2006) 2077 
  [arXiv:hep-ph/0606111].


\bibitem{otherexp8}
  P.~Huber, M.~Lindner, M.~Rolinec and W.~Winter,
  Phys.\ Rev.\  D {\bf 74} (2006) 073003 
  [arXiv:hep-ph/0606119].

\bibitem{BNLreport}
  D.~Beavis {\it et al.} [E889 Collaboration],
  Physics Design Report, BNL No. 52459 (April 1995).


\bibitem{MN97}
  H.~Minakata and H.~Nunokawa,
  Phys.\ Lett.\ B {\bf 413} (1997) 369 
  [arXiv:hep-ph/9706281].


\bibitem{silver}
  A.~Donini, D.~Meloni and P.~Migliozzi,
  Nucl.\ Phys.\ B {\bf 646} (2002) 321 
  [arXiv:hep-ph/0206034];
%
  D.~Autiero {\it et al.},
  Eur.\ Phys.\ J.\ C {\bf 33} (2004) 243 
  [arXiv:hep-ph/0305185].


\bibitem{BMW02off}
  V.~Barger, D.~Marfatia and K.~Whisnant,
  Phys.\ Rev.\ D {\bf 66} (2002) 053007 
  [arXiv:hep-ph/0206038].


\bibitem{SN}
  O.~Mena, S.~Palomares-Ruiz and S.~Pascoli,
  Phys.\ Rev.\ D {\bf 72} (2005) 053002 
  [arXiv:hep-ph/0504015];
%
  and {\it ibid.} {\bf 73} (2006) 073007 
  [arXiv:hep-ph/0510182].


\bibitem{twodetect}
  M.~Ishitsuka, T.~Kajita, H.~Minakata and H.~Nunokawa,
  Phys.\ Rev.\ D {\bf 72} (2005) 033003 
  [arXiv:hep-ph/0504026];
%
  K.~Hagiwara, N.~Okamura and K.~i.~Senda,
  Phys.\ Lett.\ B {\bf 637} (2006) 266 
  [arXiv:hep-ph/0504061];
%
  T.~Kajita, H.~Minakata, S.~Nakayama and H.~Nunokawa,
  Phys.\ Rev.\  D {\bf 75} (2007) 013006 
  [arXiv:hep-ph/0609286].


\bibitem{CDFL07}
  P.~Coloma, A.~Donini, E.~Fernandez-Martinez and J.~Lopez-Pavon,
  arXiv:0712.0796 [hep-ph].


\bibitem{atmmatter1}  
  S.~T.~Petcov, 
  Phys.\ Lett.\ B {\bf 434} (1998) 321  
  [Erratum-{\it ibid.}\ B {\bf 444} (1998) 584]
  [arXiv:hep-ph/9805262];
%
  E.~K.~Akhmedov,
  Nucl.\ Phys.\ B {\bf 538} (1999) 25 
  [arXiv:hep-ph/9805272];
%
  M.~V.~Chizhov, M.~Maris and S.~T.~Petcov, 
  hep-ph/9810501;
%
  M.~V.~Chizhov and S.T.~Petcov, 
  Phys.\ Rev.\ Lett.\ {\bf 83} (1999) 1096
  [arXiv:hep-ph/9903399];
%
  {\it ibid.} {\bf 85}, 3979 (2000); 
%
  and Phys.\ Rev.\ D {\bf 63} (2001) 073003 
  [arXiv:hep-ph/9903424].


\bibitem{mantle}
  M.~C.~Ba\~nuls, G.~Barenboim and J.~Bernab\'eu,
  Phys.\ Lett.\ B {\bf 513} (2001) 391 
  [arXiv:hep-ph/0102184];
%
  J.~Bernab\'eu and S.~Palomares-Ruiz,
  hep-ph/0112002;
%
  and Nucl.\ Phys.\ Proc.\ Suppl.\ {\bf 110} (2002) 339  
  [arXiv:hep-ph/0201090].


\bibitem{core}
  J.~Bernab\'eu, S.~Palomares-Ruiz, A.~P\'{e}rez and S.~T.~Petcov,
  Phys.\ Lett.\ B {\bf 531} (2002) 90 
  [arXiv:hep-ph/0110071];
%
  S.~Palomares-Ruiz and J.~Bernab\'eu,
  Nucl.\ Phys.\ Proc.\ Suppl.\ {\bf 138} (2005) 398 
  [arXiv:hep-ph/0312038].
 

\bibitem{atmmatter2}
  J.~Bernab\'eu, S.~Palomares Ruiz and S.~T.~Petcov,
  Nucl.\ Phys.\ B {\bf 669} (2003) 255 
  [arXiv:hep-ph/0305152];
%
  S.~Palomares-Ruiz and S.~T.~Petcov,
  Nucl.\ Phys.\ B {\bf 712} (2005) 392 
  [arXiv:hep-ph/0406096];
%
  S.~T.~Petcov and S.~Palomares-Ruiz,
  hep-ph/0406106;
%
  S.~T.~Petcov and T.~Schwetz,
  Nucl.\ Phys.\  B {\bf 740} (2006) 1
  [arXiv:hep-ph/0511277].


\bibitem{atmmatter3}
  E.~K.~Akhmedov {\it et al.},
  Nucl.\ Phys.\ B {\bf 542} (1999) 3 
  [arXiv:hep-ph/9808270];
%
  D.~Indumathi and M.~V.~N.~Murthy,
  Phys.\ Rev.\ D {\bf 71} (2005) 013001 
  [arXiv:hep-ph/0407336];
%
  R.~Gandhi {\it et al.}, 
  Phys.\ Rev.\  D {\bf 73} (2006) 053001
  [arXiv:hep-ph/0411252];
%
  D.~Indumathi, M.~V.~N.~Murthy, G.~Rajasekaran and N.~Sinha,
  Phys.\ Rev.\  D {\bf 74} (2006) 053004
  [arXiv:hep-ph/0603264];
%
  R.~Gandhi {\it et al.}, 
  arXiv:hep-ph/0506145.


\bibitem{SNastro}
  C.~Lunardini and A.~Y.~Smirnov,
  Nucl.\ Phys.\ B {\bf 616} (2001) 307  
  [arXiv:hep-ph/0106149];
%
  and JCAP {\bf 0306} (2003) 009 
  [arXiv:hep-ph/0302033];
%
  K.~Takahashi and K.~Sato,
  Phys.\ Rev.\ D {\bf 66} (2002) 033006  
  [arXiv:hep-ph/0110105];
%
  and Prog.\ Theor.\ Phys.\ {\bf 109} (2003) 919
  [arXiv:hep-ph/0205070];
%
  A.~S.~Dighe, M.~T.~Keil and G.~G.~Raffelt,
  JCAP {\bf 0306} (2003) 005 
  [arXiv:hep-ph/0303210]; 
%
  and {\it ibid.} {\bf 0306} (2003) 006
  [arXiv:hep-ph/0304150];
%
  A.~S.~Dighe, M.~Kachelriess, G.~G.~Raffelt and R.~Tomas,
  JCAP {\bf 0401} (2004) 004 
  [arXiv:hep-ph/0311172];
%
   A.~Bandyopadhyay, S.~Choubey, S.~Goswami and K.~Kar,
  hep-ph/0312315;
%
  V.~Barger, P.~Huber and D.~Marfatia,
  Phys.\ Lett.\  B {\bf 617} (2005) 167
  [arXiv:hep-ph/0501184].


\bibitem{CPT}
  P.~I.~Krastev and S.~T.~Petcov, 
  Phys.\ Lett.\ B {\bf 205} (1988) 84;
%
  J.~Arafune and J.~Sato,
  Phys.\ Rev.\ D {\bf 55} (1997) 1653
  [arXiv:hep-ph/9607437];
%
  J.~Bernab\'eu, Proc.\ WIN'99, World Scientific (2000), p. 227, 
  hep-ph/9904474;
%
  M. Freund, M. Lindner and A. Romanino,
  Nucl.\ Phys.\ B {\bf 562} (1999) 29 
  [arXiv:hep-ph/9903308];
%
  J.~Bernab\'eu and M.~C.~Ba\~nuls,
  Nucl.\ Phys.\ Proc.\ Suppl.\ {\bf 87} (2000) 315 
  [arXiv:hep-ph/0003299].


\bibitem{betabeamhigh}
  J.~Burguet-Castell {\it et al.},
  Nucl.\ Phys.\ B {\bf 695} (2004) 217 
  [arXiv:hep-ph/0312068];
%
  J.~Burguet-Castell {\it et al.},
  Nucl.\ Phys.\ B {\bf 725} (2005) 306 
  [arXiv:hep-ph/0503021].
  

\bibitem{bblindner}
  P.~Huber, M.~Lindner, M.~Rolinec and W.~Winter,
  Phys.\ Rev.\  D {\bf 73} (2006) 053002 
  [arXiv:hep-ph/0506237].


\bibitem{lowgamma}
  M.~Mezzetto,
  J.\ Phys.\ G {\bf 29} (2003) 1771 
  [arXiv:hep-ex/0302007];
%
  J.~Bouchez, M.~Lindroos and M.~Mezzetto,
  AIP Conf.\ Proc.\  {\bf 721} (2004) 37 
  [arXiv:hep-ex/0310059];
%
  M.~Mezzetto,
  Nucl.\ Phys.\ Proc.\ Suppl.\ {\bf 143} (2005) 309 
  [arXiv:hep-ex/0410083].


\bibitem{CMMS06}
  J.~E.~Campagne, M.~Maltoni, M.~Mezzetto and T.~Schwetz,
  JHEP {\bf 0704} (2007) 003 
  [arXiv:hep-ph/0603172].


\bibitem{alternating}
  A.~Donini and E.~Fernandez-Martinez,
  Phys.\ Lett.\  B {\bf 641} (2006) 432 
  [arXiv:hep-ph/0603261].


\bibitem{betaCERNupgrade}
  A.~Donini {\it et al.}, 
  arXiv:hep-ph/0511134;
%
  and Eur.\ Phys.\ J.\  C {\bf 48} (2006) 787 
  [arXiv:hep-ph/0604229];
%
  A.~Donini {\it et al.},
  arXiv:hep-ph/0703209.


\bibitem{bblhc}
  S.~K.~Agarwalla, A.~Raychaudhuri and A.~Samanta,
  Phys.\ Lett.\  B {\bf 629} (2005) 33 
  [arXiv:hep-ph/0505015]; 
%
  S.~K.~Agarwalla, S.~Choubey and A.~Raychaudhuri,
  Nucl.\ Phys.\  B {\bf 771} (2007) 1 
  [arXiv:hep-ph/0610333];
%
  S.~K.~Agarwalla, S.~Choubey, S.~Goswami and A.~Raychaudhuri,
  Phys.\ Rev.\  D {\bf 75} (2007) 097302 
  [arXiv:hep-ph/0611233];
%
  S.~K.~Agarwalla, S.~Choubey and A.~Raychaudhuri,
  arXiv:0711.1459 [hep-ph].
 

\bibitem{Boulby}
  N.~Spooner, private communication.

  
\bibitem{feasibility}
  B.~Autin {\it et al.},
  J.\ Phys.\ G {\bf 29} (2003) 1785
  [arXiv:physics/0306106].


\bibitem{rubbia}
  C.~Rubbia, A.~Ferrari, Y.~Kadi and V.~Vlachoudis,
  Nucl.\ Instrum.\ Meth.\ A {\bf 568} (2006) 475 
  [arXiv:hep-ph/0602032].


\bibitem{cern-study}
  \url{http://cern.ch/beta-beam} 


\bibitem{aC00} 
  A.~Cervera {\it et al.}, 
  Nucl.\ Phys.\  B {\bf 579} (2000) 17 
  [Erratum-ibid.\  B {\bf 593} (2001) 731 ]
  [arXiv:hep-ph/0002108].


\bibitem{lowNuFact}
  S.~Geer, O.~Mena and S.~Pascoli,
  Phys.\ Rev.\  D {\bf 75} (2007) 093001 
  [arXiv:hep-ph/0701258];
%
  A.~D.~Bross {\it et al.}, 
  arXiv:0709.3889 [hep-ph].


\bibitem{MINOSatm}
  P.~Adamson {\it et al.}  [MINOS Collaboration],
  Phys.\ Rev.\  D {\bf 75} (2007) 092003 
  [arXiv:hep-ex/0701045].


\bibitem{andy}
  A.~Blake, private communication.


\bibitem{davide}
  D.~Meloni, 
  arXiv:0802.0086 [hep-ph].


\end{thebibliography}
\end{document}